\documentclass[preprintnumbers,prd,onecolumn,floatfix,superscriptaddress, nofootinbib]{revtex4-2}
\usepackage{graphicx}
\usepackage{setspace}
\usepackage{subfig}
\usepackage{epsfig}
\usepackage{bm}
\usepackage{amssymb}
\usepackage{float}
\usepackage{amsmath}
\usepackage{dcolumn}
\usepackage[colorlinks]{hyperref}
\usepackage[usenames,dvipsnames]{color}
\usepackage{enumitem}

\hypersetup{ breaklinks=true, pdfstartview={FitH}, colorlinks=true, linkcolor=blue, citecolor=red, filecolor=magenta, urlcolor=blue, anchorcolor=green, linktocpage=true }

\begin{document}

\title{ Cosmic observation of a model in the horizon of $ f(Q, C) $-gravity}
\author{Shaily}
\email{shaily@bennett.edu.in}  
\affiliation{School of Computer Science Engineering and Technology, Bennett University, Greater Noida, India}
\author{J. K. Singh}
\email{jksingh@nsut.ac.in}
\affiliation{Department of Mathematics, Netaji Subhas University of Technology, New Delhi, India}
\author{Mohit Tyagi}
\email{mohit.tyagi@kiet.edu}  
\affiliation{KIET Group of Institutions, Delhi-NCR, Ghaziabad, India}
\author{Joao R. L. Santos}
\email{joaorafael@df.ufcg.edu.br}
\affiliation{UFCG-Universidade Federal de Campina Grande-Unidade Academica de Fisica, 58429-900 Campina Grande, PB, Brazil}

\begin{abstract}
\begin{singlespace}

In this work, we developed a cosmological model in $ f(Q, C) $ gravity within the framework of symmetric teleparallel geometry. In addition to the non-metricity scalar $Q $, our formulation includes the boundary term $ C $, which accounts for its deviation from the standard Levi-Civita Ricci scalar $ R^* $ in the Lagrangian. We derived the field equations for the metric and affine connection, employed them within a cosmological setting, and a vanishing affine connection to derive modified Friedmann equations. We used the latest observational dataset OHD in the redshift range $ z \in [0, 2.36]$, Pantheon + SH0ES in the redshift range $ z \in (0.01, 2.26)$, BAO, and the joint datasets OHD + Pantheon + SH0ES and OHD + Pantheon + SH0ES + BAO to constrain the parameters of our model by employing Markov Chain Monte Carlo (MCMC) method to minimize the $\chi^2$ term. Using the constrained free model parameters, we carefully analyzed the behavior of different physical parameters and verified that the model transits from deceleration to acceleration. Finally, we observed that the model demonstrates an expanding quintessence dark energy model and converges to the $ \Lambda $CDM in later times.

\end{singlespace}
\end{abstract}
 
\maketitle
PACS numbers: {04.20.-q, 04.50.Kd, 98.80.Es}\\
Keywords: $ f(Q, C) $ gravity, Observational analysis, Phase Transition, Perturbation analysis

\section{Introduction} \label{I}
\qquad Numerous theories of gravitational modification have been introduced to address theoretical hitches in general relativity (GR), like cosmological constant problem and non-renormalizability, and to reconcile its predictions with observational data. There are various approaches to developing these modifications. One approach is to extend the standard curvature formulation of gravity, leading to theories like $ f(R) $ gravity, $ f(G) $ gravity, Lovelock gravity, and others. Another approach is based on the torsional formulation of gravity, resulting in theories such as $f(T)$ gravity and $ f(T, T_G) $ gravity \cite{Iosifidis:2023pvz, Balhara:2023mgj, Boehmer:2021aji, BeltranJimenez:2017tkd, Ferraro:2006jd, Singh:2024ckh, Nagpal:2018mpv, Singh:2024aml, Shaily:2024rjq, Singh:2024gtz, Singh:2022ptu, Singh:2022wwa, Pawar:2024juv, Singh:2022jue, Singh:2018xjv, Singh:2022nfm, Shaily:2024nmy, Singh:2023bjx}. Based on curvature, Einstein's theory of general relativity (GR) has been remarkably successful due to its strong theoretical coherence and close alignment with experimental observations. However, this success has overshadowed the recognition of two alternative, equally valid formulations of GR in a spacetime where curvature is absent. In such contexts, gravity can be entirely explained by either torsion or the non-metricity of the spacetime. 

In the former formulation, known as the metric teleparallel theory, a metric-compatible affine connection in flat spacetime with torsion replaces the standard Levi-Civita connection, which is torsion-free and metric-compatible, upon which GR was originally formulated. This approach was pioneered by Einstein himself. In the latter case, the symmetric teleparallel theory uses an affine connection with zero curvature and torsion. Additionally, an alternative method involves using non-metricity, which gives rise to $f(Q)$ gravity. The new classes of modified gravity based on curvature, torsion, and non-metricity arise even though their unmodified theories are equivalent at the equation level.

This is a consequence of the non-metricity scalar $ Q $ and the torsion scalar $ T $ differ from the conventional Levi-Civita Ricci scalar $ R^* $\footnote{all quantities indicated with a ($*$) symbol are computed using the Levi-Civita connection $\Gamma^*$} of general relativity by a total divergence term: specifically, $ R^*=-T+B $ and $ R^*=Q+C $. As a result, the arbitrary functions $ f(R^*)$, $ f(T) $, and $f(Q)$ are no longer equivalent to a total derivative. Furthermore, scalar fields can be incorporated into these frameworks, resulting in scalar-tensor, scalar-torsion, and scalar-non-metricity theories \cite{CANTATA:2021asi, Capozziello:2011et, Heisenberg:2018vsk, Nojiri:2005jg, Bengochea:2008gz, Cai:2015emx, Kofinas:2014owa, Iosifidis:2018zjj, DAmbrosio:2021pnd, Iosifidis:2020zzp, Iosifidis:2020gth, Lu:2019hra, Anagnostopoulos:2021ydo, Solanki:2022rwu, Beh:2021wva, Maurya:2024qtu, Mandal:2020buf, Khyllep:2022spx, Lymperis:2022oyo, Dimakis:2021gby, Anagnostopoulos:2022gej, Bajardi:2020fxh, Subramaniam:2023old, Shabani:2023xfn, Shi:2023kvu, Goswami:2023knh}.

In the context of teleparallel gravities, the boundary term $B$ can be incorporated into the Lagrangian, leading to $f(T, B)$ theories, which naturally display a more complex phenomenology. In this work, we intend to explore this direction by developing $f(Q, C)$ gravity and applying it within a cosmological framework. Within the framework of metric-affine theories, symmetric teleparallel gravity and its variations are gaining prominence in the ongoing efforts to develop a unified theory of gravity. Here, ``symmetric” means that parallelograms formed by parallel transport of two vectors do not close, indicating zero torsion. On the other hand, ``teleparallel” means that a vector transported around a closed loop remains unchanged, indicating zero curvature. This method describes the geometric and dynamical effects using the non-metricity of the affine connection, characterized by zero curvature and torsion. Unlike in General Relativity, which uses the Levi-Civita connection, the metric tensor and the affine connection in this framework are treated as separate entities that interact through field equations. A major difference between $f(R^*)$ and $f(Q)$ theories is that $f(Q)$ is a second-order theory like General Relativity (GR), whereas $f(R^*)$ theory leads to fourth-order field equations. One method to increase the order in symmetric teleparallel theories is by adding higher-order terms such as $\nabla Q$ and $\nabla^k Q$ to the Lagrangian. However, a more intuitive approach is to incorporate the boundary term $C$ into the Lagrangian, as shown in studies \cite{Iosifidis:2023pvz, BeltranJimenez:2017tkd, BeltranJimenez:2019tme, Lymperis:2022oyo, Mandal:2020buf}. It is important to note that this $C$ is not the same as the Gibbons-Hawking-York boundary term, which is required for a well-defined variational formulation of GR in spacetimes with boundaries. The resulting theory is commonly referred to as the symmetric teleparallel equivalent of the $f(R^*)$ theory because with a specific choice of the function $f(Q, C)=f(Q+C)$, we can derive the $f(R^*)$ theory \cite{Farrugia:2018gyz, Paliathanasis:2017flf, Bahamonde:2015zma, Paliathanasis:2017efk, Runkla:2018xrv, Capozziello:2023vne, Bahamonde:2021gfp, DeFelice:2010nf, Jarv:2018bgs, Horndeski:1974wa, Bahamonde:2022cmz}.

The motivation behind this work is the model discussed by Avik De \textit{et al.} \cite{De:2023xua}. In that article, the authors pointed out some perspectives for further applications of their model, and some of them are addressed here, where we approached both observational and perturbation analyses. Our work is organized in the following nutshell: In Sec. \ref{sec:2}, we obtain the field equation with vanishing affine connection $ \Gamma^k_{ij} = 0 $ and choose a suitable Ansatz of scale factor for further analysis. In the next section \ref{sec:3}, we study the different observational datasets and methodologies to constrain the free model parameters. In Sec. \ref{sec:4}, we explore the physical and dynamical cosmic parameters, energy conditions, statefinder diagnostic technique, and perturbation analysis. Finally, in section \ref{sec:5}, we present our conclusions and perspectives.  

\section{$ f(Q, C) $ gravity with FLRW Cosmology} \label{sec:2}

\qquad As previously mentioned, the affine connection's incompatibility with the metric specifically, the non-vanishing covariant derivative of the metric tensor is identified by the non-metricity tensor \cite{De:2023xua, Gadbail:2023mvu}.

\begin{equation}\label{1}
Q_{kij} = \nabla_k g_{ij} = \partial_k g_{ij} - \Gamma^\beta_{ik} g_{\beta j} - \Gamma^\beta_{jk} g_{\beta i} \neq 0.
\end{equation}
Furthermore, it illustrates the gravitational interaction exclusively within these gravitational theories. The generic affine connection $\Gamma$ can always be expressed in terms of the Levi-Civita connection $\Gamma^*$ as

\begin{equation}\label{2}
\Gamma^k_{ij} = \Gamma^{*k}_{ij} + L^k_{ij},
\end{equation}
where the disformation tensor $L^k_{ij}$ is defined as

\begin{equation}\label{3}
L^k_{ij} = \frac{1}{2}(Q^k_{ij} - Q_{ij}^k - Q_{ji}^k).
\end{equation}

The superpotential tensor \(P^k_{ij}\) is written as

\begin{equation}\label{4}
P^k_{ij} = \frac{1}{4}(-2L^k_{ij} + Q^k g_{ij} - \tilde{Q}^k g_{ij} - \delta^k_{(i} Q_{j)})\,,
\end{equation}
and, the non-metricity scalar $Q$ is denoted as

\begin{equation}\label{5}
Q = Q_{\lambda\beta\gamma} P^{\lambda\beta\gamma}.
\end{equation}

Now, by adding of the constraints on torsion and curvature \textit{i.e.} $T^\lambda_{ij} := 2 \Gamma^\lambda_{[ji]} = 0 $ and $ R^k_{i \lambda j} := 2 \partial_{[\lambda} \Gamma^k_{i|j]} + 2 \Gamma^k_{\sigma[\lambda} \Gamma^\sigma_{i|j]} = 0 $, one can derive the following equations:

\begin{equation}\label{6}
R^*_{ij} + \nabla^*_\lambda L^\lambda_{ij} - \nabla^*_j \tilde{L}_i + \tilde{L}_\lambda L^\lambda_{ij} - L_{\lambda\beta j} L^{\beta \lambda}_i = 0,
\end{equation}

\begin{equation}\label{7}
R^* + \nabla^*_\lambda (L^\lambda - \tilde{L}^\lambda) - Q = 0.
\end{equation}

Once \( Q^\lambda - \tilde{Q}^\lambda = L^\lambda - \tilde{L}^\lambda \), following from the earlier statement, the boundary term is also defined as

\begin{equation}\label{8}
C = R^* - Q = -\nabla^*_{\lambda} (Q^\lambda - \tilde{Q}^\lambda) = -\frac{1}{\sqrt{-g}} \partial_\lambda ( \sqrt{-g} (Q^\lambda - \tilde{Q}^\lambda)).
\end{equation}

Considering the previous discussion, $ f(Q, C) $ gravity combines elements from both $f(R^*)$ and $f(Q)$ theories within a unified framework. The action is specified by

\begin{equation}\label{9}
S = \int \left( \frac{1}{2\kappa} f(Q, C) + \mathcal{L}_m \right) \sqrt{-g} \, d^4x,
\end{equation}
where $f(Q, C)$ is an arbitrary function of $Q$ and $C$ and $\mathcal{L}_m$ is a matter Lagrangian. Now, variation in action concerning metric tensor yields the field equation as

\begin{equation}\label{10}
\kappa T_{ij} = -\frac{f}{2} g_{ij} + 2P^k_{ij} \nabla_k (f_Q - f_C) + \left( G^*_{ij} + \frac{Q}{2} g_{ij} \right) f_Q + \left( \frac{C}{2} g_{ij} - \nabla^*_i \nabla^*_j + g_{ij} \nabla^{*\lambda} \nabla^*_\lambda \right) f_C.
\end{equation}

Also, the effective stress-energy tensor is defined as
\begin{equation}\label{11}
T^\text{eff}_{ij} = T_{ij} + \frac{1}{\kappa} \left[ \frac{f}{2} g_{ij} - 2 P^k_{ij} \nabla_k (f_Q - f_C) - \frac{Q f_Q}{2} g_{ij} - \left( \frac{C}{2} g_{ij} - \nabla^*_i \nabla^*_j + g_{ij} \nabla^{*\lambda} \nabla^*_\lambda \right) fC \right],
\end{equation}
thus we have
\begin{equation}\label{12}
G^\prime_{ij} = \kappa f_Q T^\text{eff}_{ij}.
\end{equation}

Here, $ f_C $ and $f_Q$ are the derivatives of $f$ with respect to $C$ and $Q$, respectively. It is crucial to note that unlike in General Relativity, the teleparallel theory features an affine connection separate from the metric tensor, both of which are treated as dynamic variables. Consequently, varying the action concerning the affine connection yields the connection field equation $ (\nabla_i - \tilde{L}_i)(\nabla_j - \tilde{L}_j) [4(f_Q - f_C) P^{ij}_k] = 0$, in the absence of the hyper-momentum tensor.

We explore $f(Q, C)$ gravity within a cosmological context, specifically focusing on $f(Q, C)$ gravity. We investigate a homogeneous and isotropic flat FLRW spacetime described by the line element.

\begin{equation}\label{13}
ds^2 = -dt^2 + a^2(t)\sum_{n=1}^{3}dx_n^2,
\end{equation}
where $a(t)$ is the scale factor of the Universe. Within the framework of $f(Q, C)$ gravity, we obtain the effective dark-energy sector by considering the energy-momentum tensor as
\begin{equation}\label{14}
T^\text{DE}_{ij} = \frac{1}{f_Q} \left[ \frac{f}{2} g_{ij} - 2 P^k_{ij} \nabla_k (f_Q - f_C) - \frac{Q f_Q}{2} g_{ij} - \left( \frac{C}{2} g_{ij} - \nabla^*_i \nabla^*_j + g_{ij} \nabla^{*\lambda} \nabla^*_\lambda \right) f_C \right].
\end{equation}

The cosmological uniformity and isotropy of the FLRW metric (\ref{13}) are exemplified by spatial rotations and translations. An asymmetric teleparallel affine connection, free from torsion and curvature, exhibits both spherical and translational symmetries. This implies that the derivative operations of the connection coefficients concerning the vector fields generating spatial rotations and translations vanish. Three specific types of affine connections demonstrate these symmetries. We consider the case where the affine connection vanishes, $ \Gamma^k_{ij} = 0 $, and when fixing the coincident gauge, we obtained the Friedmann-like equations as \cite{Shi:2023kvu, Hohmann:2021ast}

\begin{equation}\label{15}
3H^2 = \kappa (\rho_m + \rho_\text{DE}), 
\end{equation}
\begin{equation}\label{16}
- (2\dot{H} + 3H^2) = \kappa (p_m + p_\text{DE}),
\end{equation}
where, $\rho_m$ is the matter energy density and $p_m$ denotes the matter pressure. We introduce the effective dark-energy density and pressure as 

\begin{equation}\label{17}
\rho_{\text{DE}} = \frac{1}{\kappa} \left[ 3H^2 (1 - 2f_Q) - \frac{f}{2} + (9H^2 + 3\dot{H}) f_C - 3 H \dot{f_C} \right] 
\end{equation}

\begin{equation}\label{18}
p_{\text{DE}} = \frac{1}{\kappa} \left[ -2\dot{H} (1 - f_Q) - 3H^2 (1 - 2f_Q) + \frac{f}{2} + 2 H \dot{f_Q} - (9H^2 + 3\dot{H}) f_C + \ddot{f}_C \right],
\end{equation}
for $ ~Q = -6H^2,~R^* = 6(2H^2 + \dot{H}),~ C = R^* - Q = 6(3H^2 + \dot{H})$. \\

We observed that the equations (\ref{17}) and (\ref{18}) are reliant on the function $f(Q, C)$ and its derivatives. One might consider selecting functions that yield GR equations of motion when additional terms vanish to incorporate GR solutions alongside effects applicable during early or late times. A possible choice is the function $f(Q, C) = Q^\mu + \zeta C^\nu$, where $\mu$, $\nu$ and $\zeta$ are constants. For simplicity, we set $\mu=2$, $\nu=1$, and $\zeta=2$ and explore the behavior of dynamical and physical parameters. It is relevant to highlight that such choices generalize this model in respect to general relativity, which can be recovered by setting $\zeta=0$ and $\mu=1$.

Now, for further analysis, we need to solve the field equations. But, once there are three unknowns in equations (\ref{17}) and (\ref{18}), namely, $H$, $\rho_{\text{DE}}$ and $p_{\text{DE}}$, we need one additional equation to examine the behavior the model.
In this work, we restrict the scale factor in the shape of the hybrid expansion law. This form triggers a transition in the Universe from a decelerating phase to an accelerating phase, and it is defined as

\begin{equation}\label{19}
    a(t)=\left( \frac{t}{t_0} \right)^{\alpha} e^{\beta{\left( \frac{t}{t_0}-1\right)}},
\end{equation}
where $\alpha$ and $\beta$ are the positive constants and $t_0$ denotes the present cosmic time. In the literature, it is recognized that this formulation encompasses both power-law cosmology and exponential law cosmology as special cases. When $\alpha=0$, the hybrid expansion law in (\ref{19}) simplifies to the exponential law and when $\beta=0$, it reduces to the power-law \cite{Singh:2022eun, Koussour:2022jss}.\\

Now, the Hubble parameter $H$ can be obtained by rewriting $H=\frac{\Dot{a}}{a}$ as
\begin{equation}\label{20}
    H(t)=\frac{\alpha}{t}+\frac{\beta}{t_0}.
\end{equation}

Here, we constrain the values of $\alpha$ and $\beta$ using the observational datasets and for that purpose, we need to have the Hubble parameter as a function of the redshift ($ z $). Thus, with the relation $a=\frac{a_0}{1+z}$ (at present $a_0=1$) and equation (\ref{19}), one can obtain the value of $t$ in terms of $z$
\begin{equation}\label{21}
    t(z)=\frac{\alpha t_0}{\beta} W \left( \frac{\beta e^\frac{\beta}{\alpha}}{\alpha (1+z)^\frac{1}{\alpha}}\right),~ \text{where}~~t_0=\frac{\alpha+\beta}{H_0}
\end{equation}
therefore, we find that
\begin{equation}\label{22}
    H(z)=\frac{H_0 \beta}{\alpha+\beta} \left(\frac{1}{W \left( \frac{\beta e^\frac{\beta}{\alpha}}{\alpha (1+z)^\frac{1}{\alpha}}\right)}+1\right),
\end{equation}
where $ W $ is the Lambert function, $ H_0 $ is the present value of the Hubble parameter which will also be constrained by observational datasets.
In the next section, we use three observational datasets to find the optimal values of the free model parameters $ H_0 $, $ \alpha $, and $ \beta $. Markov Chain Monte Carlo (MCMC) method is applied for this purpose, and contours were plotted using the emcee library in Python.

\section{Observational datasets and Methodology} \label{sec:3}
\qquad Recent advances in observational cosmology have significantly deepened our comprehension of cosmic evolution, spanning from ancient origins to contemporary developments. These advancements have elucidated the characteristics of dark matter and dark energy, as well as the formation of cosmic structures. Over the past three decades, observational studies in cosmology, driven by instruments such as the Hubble Space Telescope, have generated a vast array of data. Key datasets include the Sloan Digital Sky Survey, which charts the distribution of galaxies and captures current cosmic variations; cosmic microwave background radiation, validating the Big Bang theory; quasars, providing insights into intervening matter between observers and these distant phenomena; and baryon acoustic oscillations, offering measurements of large-scale structures crucial for understanding dark energy. Additionally, supernovae type Ia, known as standard candles, is pivotal in determining cosmic distances. This study employs error bar plots of Hubble and SNeIa datasets to compare our model with the $\Lambda$CDM model. Using optimization techniques in Python, we constrain model parameters and predict the global minimum of the Hubble function as described in Eq. (\ref{22}). Noticeable fluctuations within the diagonal elements of the covariance matrix related to these parameters are observed. By utilizing these measurements as averages and applying a Gaussian prior with a fixed dispersion value $\sigma=1$, we use the emcee module of Python for mathematical research and numerical analysis. This methodology allows us to explore the parameter space surrounding local minima. Results are depicted as two-dimensional contour plots indicating $1\sigma$ and $2\sigma$ uncertainties. I've shared the details on the methodologies applied to these datasets below.

\subsection{Hubble dataset}
\qquad The Hubble parameter $H$ is a fundamental cosmic parameter that describes the expansion rate within the Universe and encapsulates crucial details about its evolutionary history. It can be defined in terms of observable quantities such as redshift, length, and time through the relation $H(z) = \frac{-1}{1+z} \frac{dz}{dt}$. Observations of $H(z)$ offer valuable insights into significant epochs of cosmic evolution, shedding light on dark matter and energy phenomena. In this study, we utilize an updated dataset consisting of $77$ data points spanning the redshift range $ z \in [0, 2.36]$  \cite{Shaily:2022enj, Singh:2023ryd, Singh:2023gxd, Singh:2024kez, Sharov:2018yvz, Stern:2009ep, Moresco:2012jh, Moresco:2016mzx, Ratsimbazafy:2017vga, BOSS:2016zkm, BOSS:2016wmc, Chuang:2013hya, BOSS:2013rlg, BOSS:2017fdr, Pan-STARRS1:2017jku}. The observational constraints on model parameters $H_0$, $\alpha$ and $\beta$, representing the most probable scenarios, are derived by minimizing the chi-square value $\chi^2_{\text{min}}$, which corresponds to maximizing the likelihood analysis. The likelihood function $\chi^2_H(H_0, \alpha, \beta) $ can be computed as follows

\begin{figure}\centering
	\subfloat[]{\includegraphics[scale=0.7]{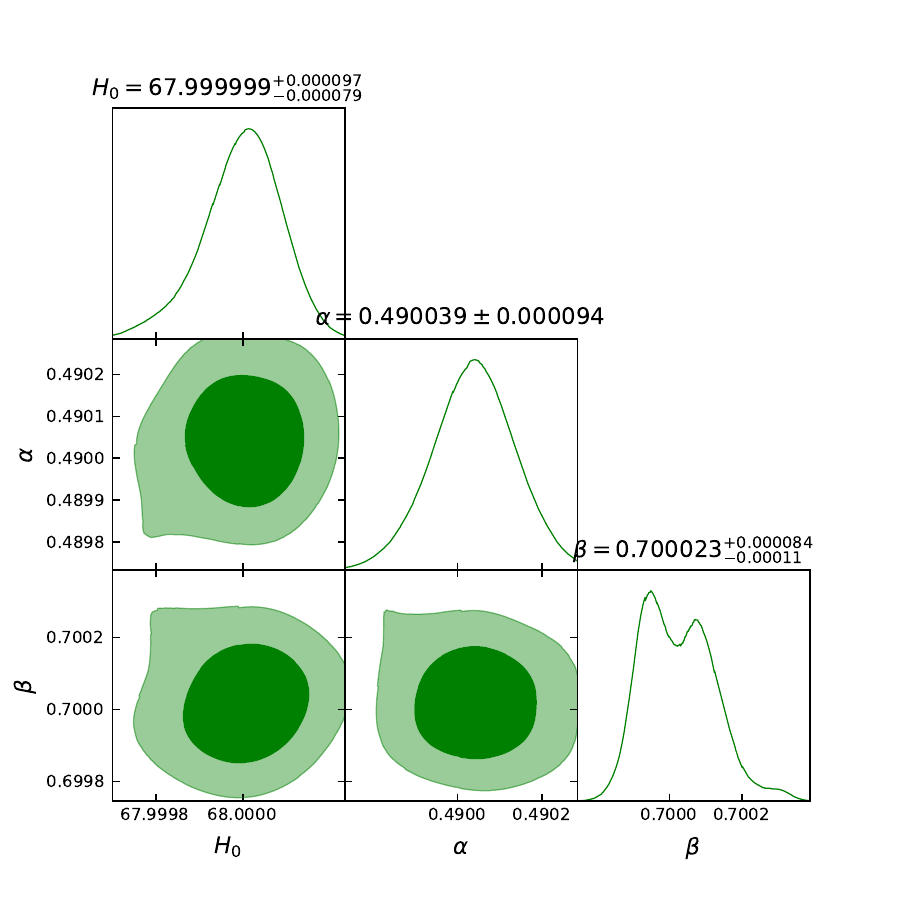}}
\caption{The posterior distribution of the reconstructed deceptive $ f(Q, C) $ model for $ 1\sigma $ and $ 2\sigma $ confidence levels (CLs) obtained from the $H(z)$ dataset.}
\label{contour1}
\end{figure}

\begin{equation}\label{23}
    \chi^2_H(H_0, \alpha, \beta) = \sum_{i=1}^{77} \frac{\left[ H_{\text{th}}(z_i, H_0, \alpha, \beta) - H_{\text{obs}}(z_i) \right]^2}{\sigma^2_H(z_i)},
\end{equation}
where $H_{\text{obs}}(z_i)$ and $H_{\text{th}}(z_i, H_0, \alpha, \beta)$ stands for the observed and theoretical values of $H$ at redshift $z_i$, respectively. $\sigma_H(z_i)$ represents the standard error in the observed value of $H$.

\subsection{Pantheon + SH0ES Dataset}
As technology progresses, our exploration of observational cosmology relies heavily on diverse datasets that illuminate various aspects, such as early cosmic evolution, the formation of structures, and the enigmatic properties of the dark Universe, particularly in explaining the accelerated expansion observed in the cosmos using cosmic mechanisms and ray detectors. Among these datasets, the Pantheon + SH0ES sample holds significant prominence, comprising $1701$ data points and serving as a leading collection of observations of supernovae type Ia. The Pantheon + SH0ES dataset spans the spectroscopically observed range of redshifts $ z \in (0.01, 2.26)$ \cite{Pan-STARRS1:2017jku, Riess:1998dv, Jha:2005jg, Hicken:2009df, Contreras:2009nt, SDSS:2014irn}. The distance moduli are calculated using the equation $\mu_{\text{th}, i} = \mu(D_L) = m - M = 5 \log_{10}(D_L) + \mu_0$, where $M$ denotes absolute magnitude, $m$ denotes apparent magnitude, and $\mu_0$ is a marginalized nuisance parameter. The parameter $\mu_0$ can be obtained as $\mu_0 = 5 \log_{10} \left( \frac{c}{H_0} \text{Mpc} \right) + 25$.

\begin{figure}\centering
	\subfloat[]{\includegraphics[scale=0.7]{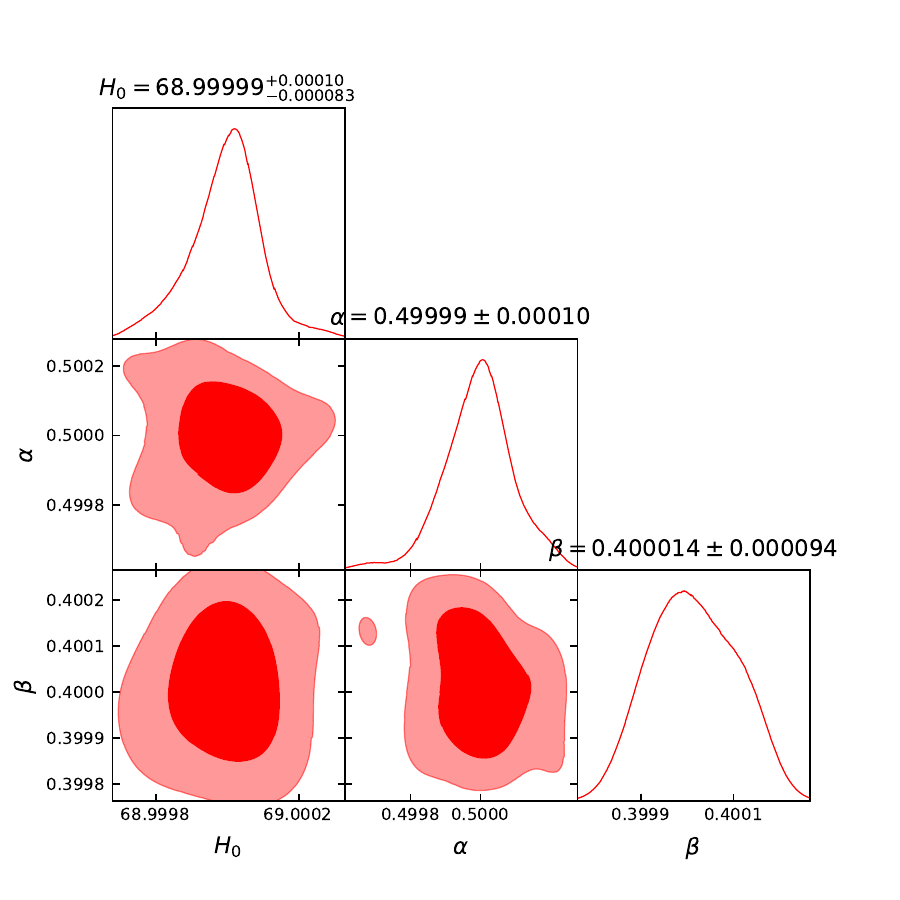}}
\caption{ The posterior distribution of the reconstructed deceptive $ f(Q, C) $ model for $ 1\sigma $ and $ 2\sigma $ CLs obtained from the Pantheon + SH0ES dataset.}
\label{contour2}
\end{figure}

To determine the optimal model parameters $H_0$, $\alpha$, and $\beta$, the theoretical distance modulus $\mu_{\text{th}, i}$ is compared with the observed distance modulus $\mu_{\text{obs}, i}$. The luminosity distance $D_L(z)$ is adjusted using the formula
\begin{equation}\label{24}
    D_L(z) = c (1 + z) \int_0^z \frac{dz'}{H(z')},
\end{equation}
where $c$ is the speed of light. The agreement between the Pantheon + SH0ES data and the model predictions is quantified by taking the $\chi^2$ function as
\begin{equation}\label{25}
    \chi^2_{\text{PN}}(H_0, \alpha, \beta) = \sum_{i=1}^{1701} \frac{[\mu_{\text{th}}(H_0, \alpha, \beta) - \mu_{\text{obs}}(z_i)]^2}{\sigma^2_{\mu}(z_i)},
\end{equation}
here $\sigma_{\mu}(z_i)$ denotes the standard error associated with the observed distance modulus $\mu_{\text{obs}}(z_i)$.

\subsection{BAO Dataset}
The exploration of baryon acoustic oscillations occurs in the early Universe, where baryons and photons are tightly coupled due to Thomson scattering. Because of the intense photon pressure, both baryons and photons behave as a unified fluid that resists gravitational collapse. Instead, they undergo oscillations, leading to the phenomenon known as Baryon Acoustic Oscillations (BAO). BAO involves analyzing the spatial distribution of galaxies to measure how cosmic structures evolve during the expansion of the cosmos \cite{Weiland:2010ij, SDSS:2009ocz, Giostri:2012ek, Blake:2011en, SDSS:2005xqv, Beutler:2011hx}. This distinction provides a means to differentiate between various types of Dark Energy (DE) models. The sound horizon $r_s$ at the epoch of photon decoupling $z^*$, which defines the characteristic scale of BAO, is given by:
\begin{equation}\label{26}
    r_s(z^*) = \frac{c}{\sqrt{3}} \int_{\frac{1}{1+z^*}}^1 \frac{da}{a^2 H(a)} \sqrt{1 + \frac{3\Omega_{b0}}{4\Omega_{\gamma0}} a},
\end{equation}
where $\Omega_{b0}$ denotes the current density of baryons and $\Omega_{\gamma0}$ stands for the current density of photons, respectively. BAO measurements utilize the following relationships:
\begin{figure}\centering
	\subfloat[]{\includegraphics[scale=0.7]{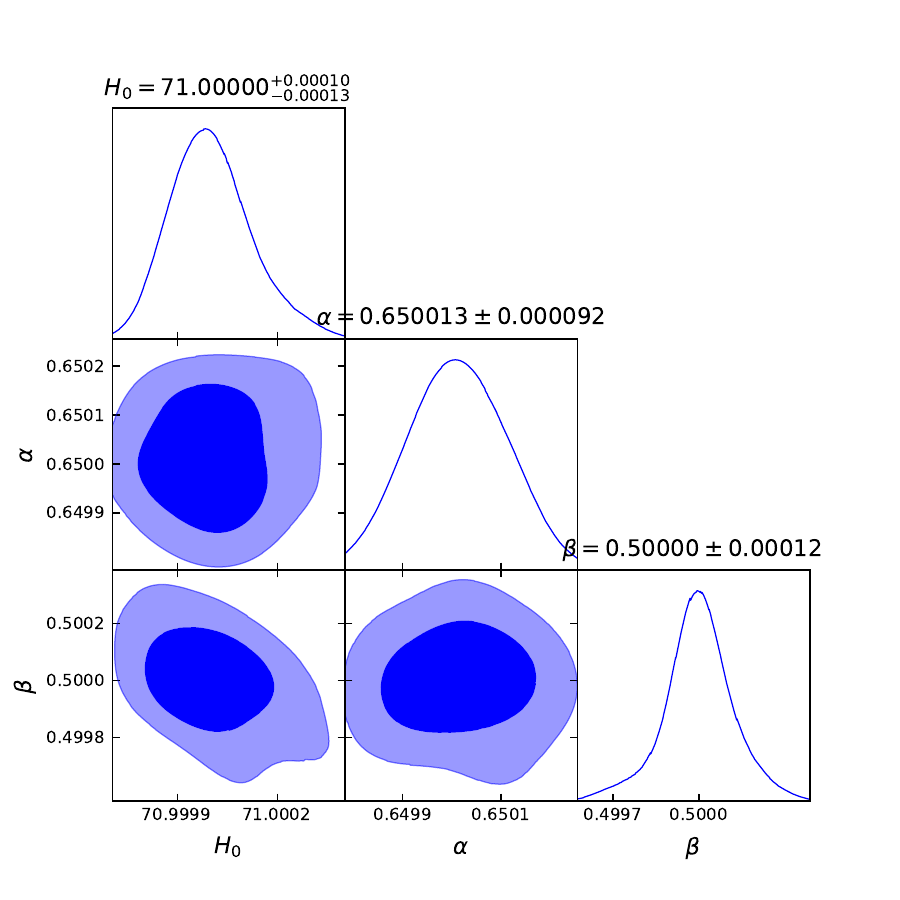}}\hfill
\caption{The posterior distribution of the reconstructed deceptive $ f(Q, C) $ model for $ 1\sigma $ and $ 2\sigma $ CLs obtained from the Pantheon + SH0ES dataset.}
\label{contour3}
\end{figure}

\begin{equation}\label{27}
    \Delta\theta = \frac{r_s}{d_A(z)}, \quad d_A(z) = \int_0^z \frac{dz'}{H(z')},
\end{equation}

\begin{equation}\label{28}
    \Delta z = H(z) \cdot r_s.
\end{equation}
Here, $\Delta z$, $\Delta\theta$ represent the observed redshift separation and the angular separation of BAO features in the $2$-point correlation function of the galaxy distribution across the sky, $d_A$ is the measured angular diameter distance. In this study, the redshift at the photon decoupling epoch is assumed to be $z^* \approx 1091$. The comoving angular diameter distance $d_A(z)$ and the dilation scale $D_V(z) = \left( \frac{d_A(z)}{z} \frac{c}{H(z)} \right)^{1/3}$ are employed in the BAO dataset of six locations for calculations involving $\frac{d_A(z^*)}{D_V(z_{\text{BAO}})}$. For the BAO dataset, the chi-square function is assumed as
\begin{equation}\label{29}
    \chi^2_{\text{BAO}} = X^T C_{\text{BAO}}^{-1} X,
\end{equation}
where $\mathbf{X}$ depends on the specific survey used, and $\mathbf{C}_{\text{BAO}}^{-1}$ denotes the inverse covariance matrix.

\begin{figure}\centering
	\subfloat[]{\includegraphics[scale=0.7]{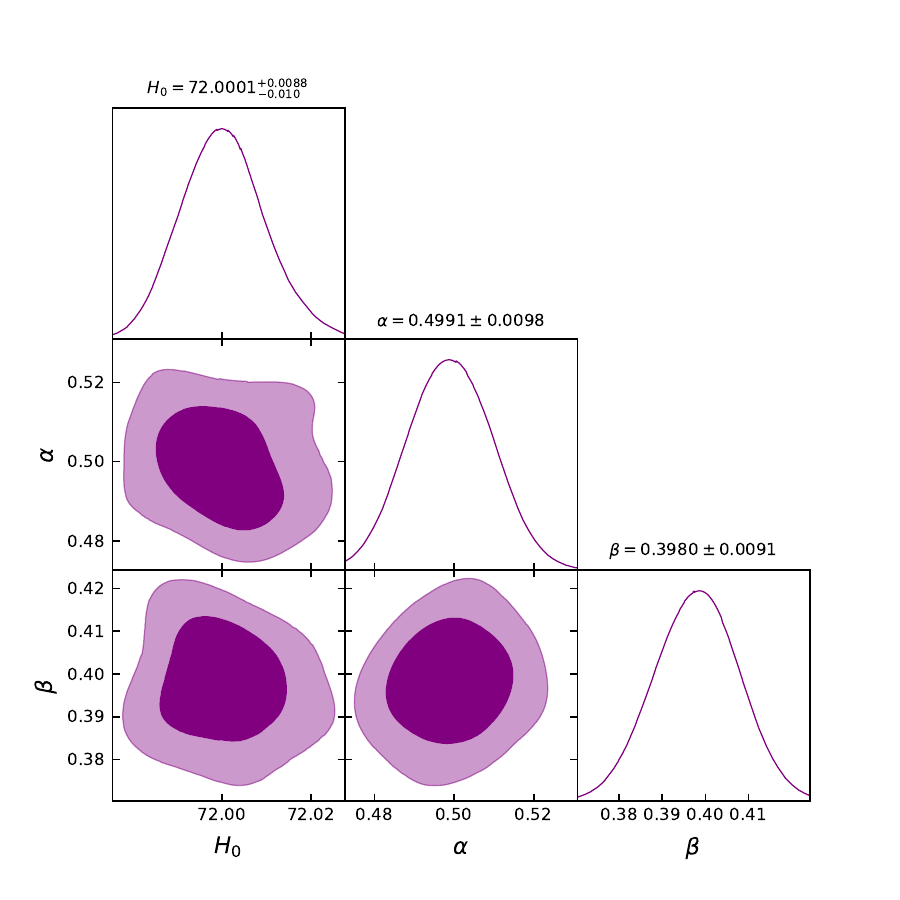}}\hfill
\caption{ The posterior distribution of the reconstructed deceptive $ f(Q, C) $ model for $ 1\sigma $ and $ 2\sigma $ CLs obtained from the $H(z)$ + Pantheon + SH0ES dataset.}
\label{contour4}
\end{figure}

\begin{figure}\centering 
	\subfloat[]{\label{erh}\includegraphics[scale=0.42]{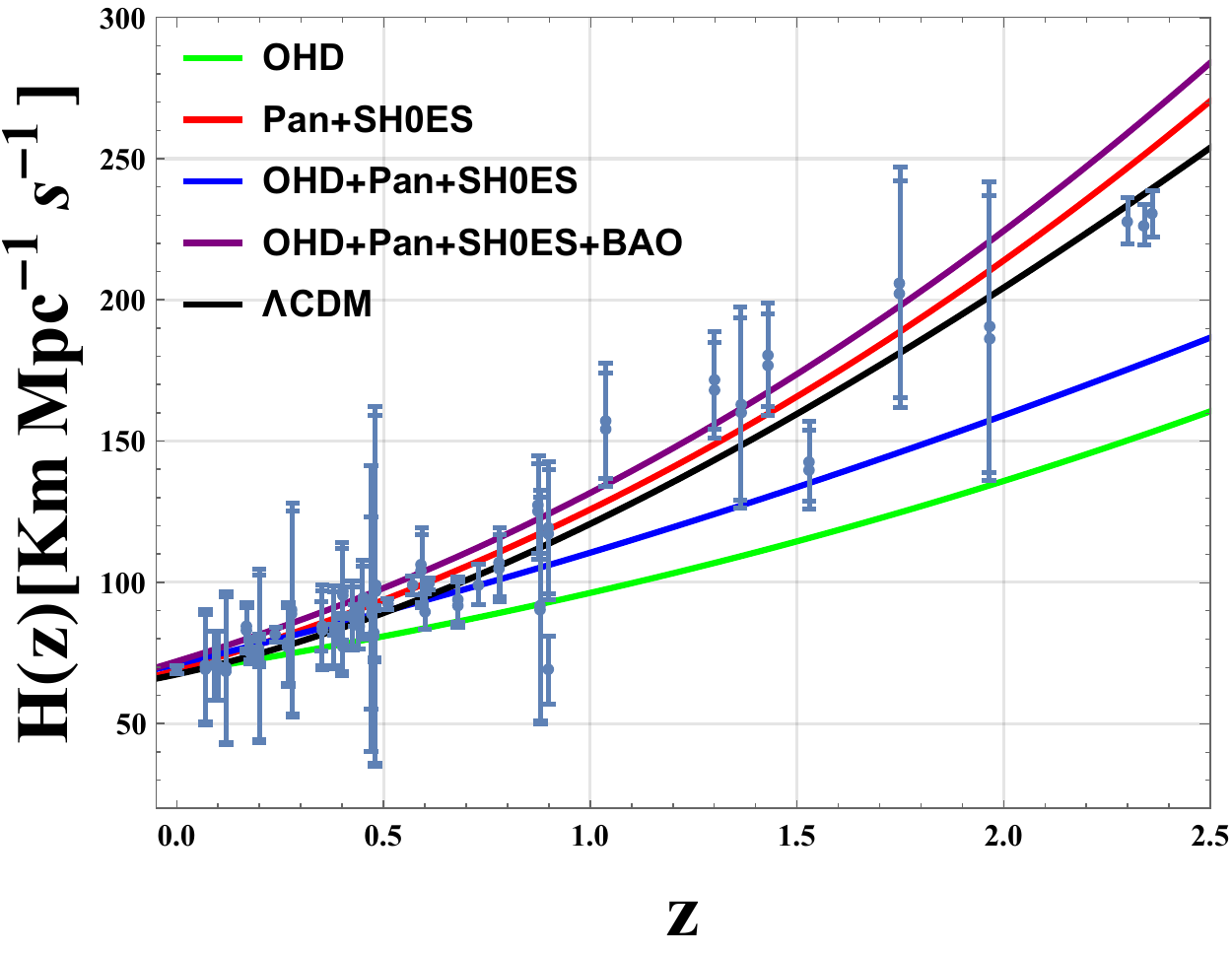}}\hfill
	\subfloat[]{\label{erm}\includegraphics[scale=0.56]{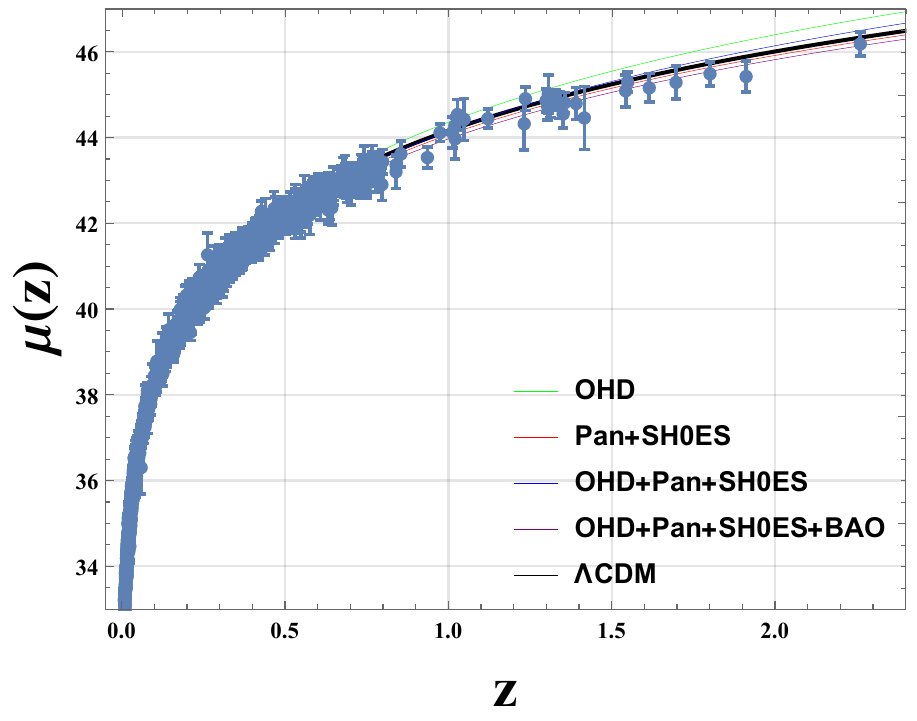}} 
\caption{ The alikeness of our model with $ \Lambda $CDM for $H(z)$ and Pantheon + SH0ES in the Error Bar plots. }
 \label{errorbar}
\end{figure}

\subsection{$H(z)$ + Pantheon + SH0ES and $H(z)$ + Pantheon + SH0ES + BAO Datasets}

\qquad The combined $\chi^2$ function for the $H(z)$ + Pantheon + SH0ES + BAO dataset is given by: $\chi^2_{\text{HP}} = \chi^2_{\text{H}} + \chi^2_{\text{PN}}$ and  $\chi^2_{\text{tot}} = \chi^2_{\text{H}} + \chi^2_{\text{PN}}+ \chi^2_{\text{BAO}}$. We establish constraints on the model parameters by employing Markov Chain Monte Carlo (MCMC) to minimize each $\chi^2$ term individually. In figs. \ref{contour1}, \ref{contour2}, \ref{contour3} and \ref{contour4}, the best fitted values of model parameters are obtained and have been summarized in Table \ref{tab1}. Also, to compare our model with $\Lambda$CDM model we have plotted the error bar fit for $H(z)$ dataset and SNeIa dataset in Fig. \ref{errorbar}. 

\begin{table}
	\caption{ The Best-fit values of the model parameters }
	\begin{center}
		\label{tab1}
		\begin{tabular}{l c c c c r} 
			\hline\hline
			\\
			{Model parameters} &  ~~~~~~~   $H(z)$ & ~~~~~   Pan+SH0ES  & ~~~~~~    $H(z)$+Pan+SH0ES & ~~~~~~    $H(z)$+Pan+SH0ES+BAO\footnote{ transition occurs from deceleration to acceleration stated in Fig. \ref{q}}\footnote{ The model represents the quintessence dark energy model at present.}\footnote{ converging to $\Lambda$CDM model in late times. } 
			\\
			\\
			\hline
			\\      
			{ $ H_0 $ }  &   ~~~~~~~ $ 67.99999_{-0.000079}^{+0.000097} $   &  ~~~~~ $ 68.99999_{-0.000083}^{+0.0001} $ & ~~~~~ $ 71.00000_{-0.00013}^{+0.0001} $ &  ~~~~~~~ $72.0001_{-0.010}^{+0.0088}$  
			\\
			\\
			{ $ \alpha $ }    & ~~~~~~~ $ 0.49004_{-0.000094}^{+0.000094} $   &  ~~~~~ $ 0.49999_{-0.0001}^{+0.0001} $ &  ~~~~~ $ 0.65001_{-0.000092}^{+0.000092} $  & ~~~~~$ 0.4991_{+0.0098}^{-0.0098} $  
			\\
			\\
			{ $\beta $ }  &  ~~~~~~~$ 0.70002_{-0.00011}^{+0.000084}  $  & ~~~~~ $  0.40001_{-0.000094}^{+0.000094} $ &  ~~~~~ $ 0.5_{-0.00012}^{+0.00012} $ & ~~~~~ $ 0.3980_{+0.0091}^{-0.0091} $ 
			\\
			\\
            { $ q_0 $ }  &  ~~~~~~~$ -0.654  $  & ~~~~~ $  -0.383 $ &  ~~~~~ $ -0.509$ & ~~~~~ $ -0.380$
            \\
            \\
            { $z_{tr} $ }  &  ~~~~~~~$ 1.97  $  & ~~~~~ $  0.71 $ &  ~~~~~ $ 1.99 $ & ~~~~~ $ 0.71 $ 
            \\
            \\
			\hline\hline  
		\end{tabular}    
	\end{center}
\end{table}

\begin{figure}\centering
	\subfloat[]{\label{q}\includegraphics[scale=0.51]{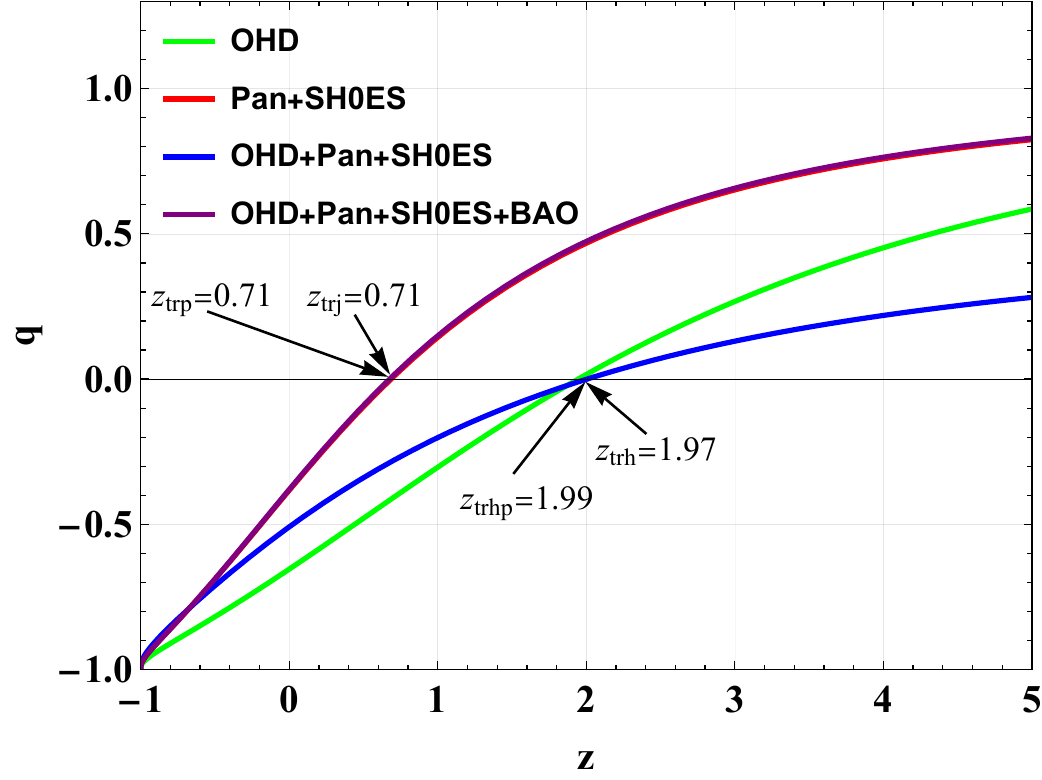}}\hfill
	\subfloat[]{\label{j}\includegraphics[scale=0.61]{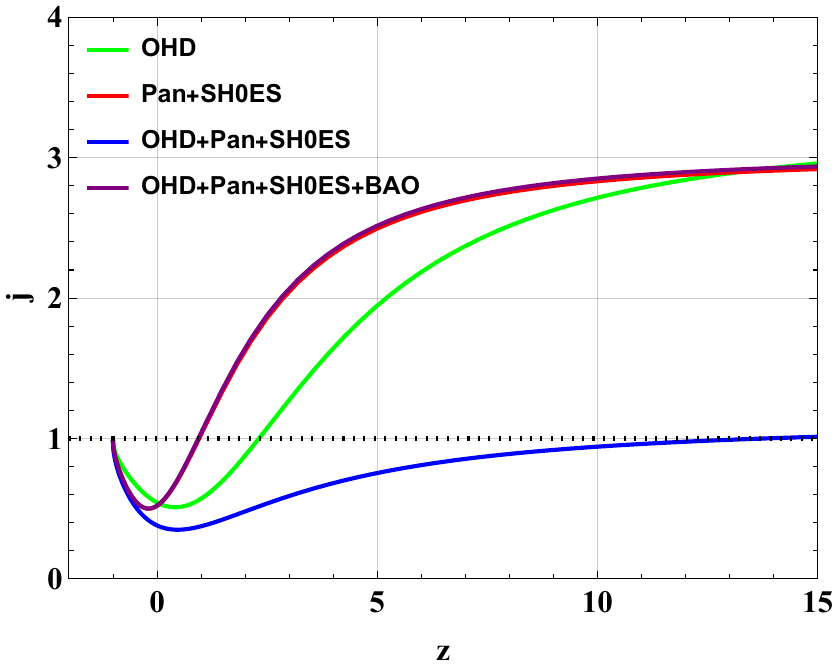}}
\caption{The evolution of the deceleration parameter ($ q $) and the jerk parameter ($ j $) over the redshift ($ z $).}
 \label{fig:3}
\end{figure}

\begin{table}
\caption{\bf Summary of the best-fit values for the Hubble parameter $ H_0 $ and the density parameter $ \Omega_{m0} $ of our model and the $\Lambda$CDM in $ f(Q, C) $ gravity in comparison with the exponential $ F(R) $ gravity with logarithmic corrections, the standard exponential $ F(R) $ model, the $ F(R) $ model, and the $\Lambda$CDM model of Odintsov et al. \cite{Odintsov:2023cli, Odintsov:2024lid}.}

\begin{center}
\label{tabparm3}
\begin{tabular}{l l l l c c} 
\hline\hline
     \\ 
  {\bf Model}  &  ~~~~~{\bf Datasets}~~~~~~~~~~~~~~~~& ~~~~ $ \boldsymbol{H_0} $~~~~~~~~~~~~~~~~& ~~~~$ \boldsymbol{\Omega_{m0}} $
        \\
        \\
        \hline
        \\  
   $ f(Q, C) $   &  ~~~~~$ H(z) $+Pan+SH0ES+BAO &~~~ $71.999_{-0.010}^{+0.010}$ &~~~$~~~~-$
      \\
       \\
           
  $\Lambda$CDM & ~~~~~$ H(z) $+Pan+SH0ES+BAO ~~~~ &~~~ $68.9989_{-0.0098}^{+0.0081}$  &~~~$0.2499_{-0.0076}^{+0.0086}$
       \\
       \\
         \hline
        \\  
   Exp $+\log F(R) $ & ~~~~~CC H(z)+SNeIa+CMB+BAO ~~~~ &~~~ $ 68.92^{+1.63}_{-1.72} $ &~~~$0.2984^{+0.0064}_{-0.0057}$  \cite{Odintsov:2024lid}
       \\
       \\
   Exp~$ +\log F(R)~ + $~axion  & ~~~~~CC H(z)+SNeIa+CMB+BAO ~~~~ &~~~ $ 69.0^{+1.72}_{-1.71} $ &~~~$0.2980^{+0.0065}_{-0.0065}$ \cite{ Odintsov:2024lid}
  \\
  \\
  Exp $ F(R)  $ & ~~~~~CC H(z)+SNeIa+CMB+BAO ~~~~ &~~~ $ 68.84^{+1.75}_{-1.64} $ &~~~$0.2913^{+0.0035}_{-0.0015}$ \cite{Odintsov:2024lid}
  \\
  \\
        \hline
        \\  
  $  F(R)~+ $ EDE   & ~~~~~CC H(z)+SNeIa+CMB+BAO ~~~~ &~~~ $ 68.93^{+1.61}_{-1.57} $ &~~~$0.294^{+0.0048}_{-0.0036}$ \cite{Odintsov:2023cli}
  \\
    \\
 $\Lambda$CDM ~~~~     & ~~~~~CC H(z)+SNeIa+CMB+BAO  & ~~~~$ 68.98^{+1.58}_{-1.60} $ &~~~$0.2908^{+0.0013}_{-0.0012}$ \cite{Odintsov:2023cli, Odintsov:2024lid} 
       \\
       \\
\hline\hline  
\end{tabular}   
\end{center}
\end{table}

\section{Dynamical and Physical behavior of the model}\label{sec:4}
\subsection{Cosmological Parameters}

\qquad In the evolution of the Universe, we have discussed various physical and kinematic parameters whose characteristics can be explored either by analyzing their formulas or interpreting graphical representations. In this section, we have examined the formulae of some essential parameters, for instance, the deceleration parameter ($ q $), jerk parameter ($ j $), and EoS parameter ($ \omega $). The deceleration parameter can be computed by the formula $q=-1-\frac{\Dot{H}}{H}$. Using the Eq. (\ref{20}), $ q $ can is given by
\begin{equation}\label{30}
    q=\frac{\alpha }{t^2 \left(\frac{\alpha }{t}+\frac{\beta }{\text{t0}}\right)^2}-1.
\end{equation}

The above expression can also be obtained in terms of redshift ($z$) using eq. (\ref{21}) as 

\begin{equation}\label{31}
    q=\frac{1}{\alpha  \left(W\left(\frac{\beta  e^{\beta /\alpha } (z+1)^{-1/\alpha }}{\alpha }\right)+1\right)^2}-1
\end{equation}

Graphical analysis is the most effective way to study the evolution of the deceleration parameter. In Fig. \ref{q}, we plot the trajectories of $ q $ using the latest observational datasets. In fig. \ref{q}, we observe that the model transits from the early decelerating to the late-times accelerating state. Such a behavior is consistent with the standard cosmology. Also, from the deceleration parameter, we can find another kinematic parameter named jerk parameter $ j $, whose equation is

\begin{equation}\label{32}
    j=q+2q^2+(1+z)\frac{dq}{dz}=1+\frac{-3 \alpha -3 \alpha  W\left(\frac{\beta  e^{\beta /\alpha } (z+1)^{-1/\alpha }}{\alpha }\right)+2}{\alpha ^2 \left(W\left(\frac{\beta  e^{\beta /\alpha } (z+1)^{-1/\alpha }}{\alpha }\right)+1\right)^3}.
\end{equation}

\begin{figure}\centering
	\subfloat[]{\label{rh}\includegraphics[scale=0.64]{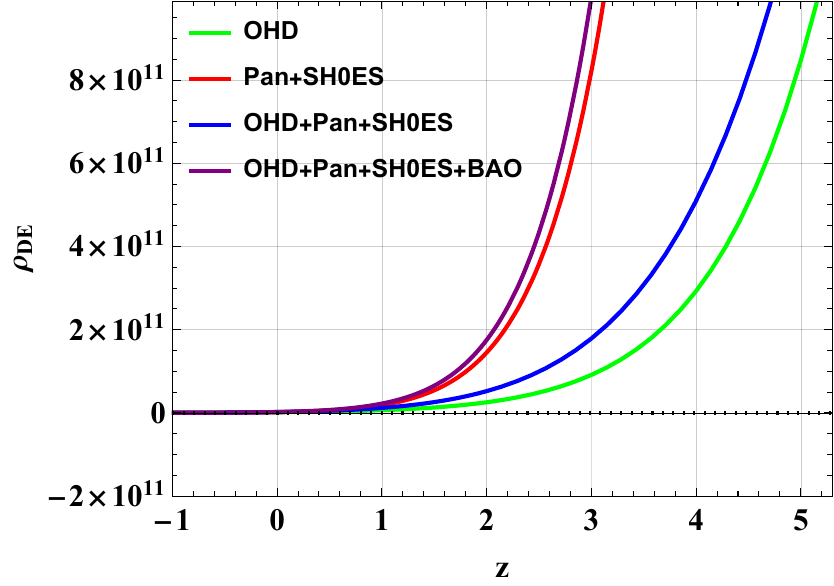}}\hfill
	\subfloat[]{\label{p}\includegraphics[scale=0.62]{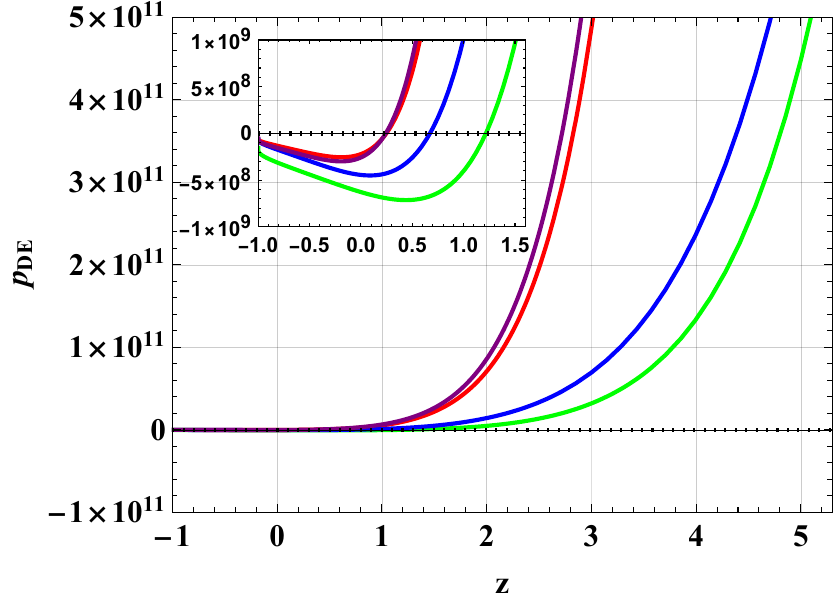}}\par
	\subfloat[]{\label{w}\includegraphics[scale=0.62]{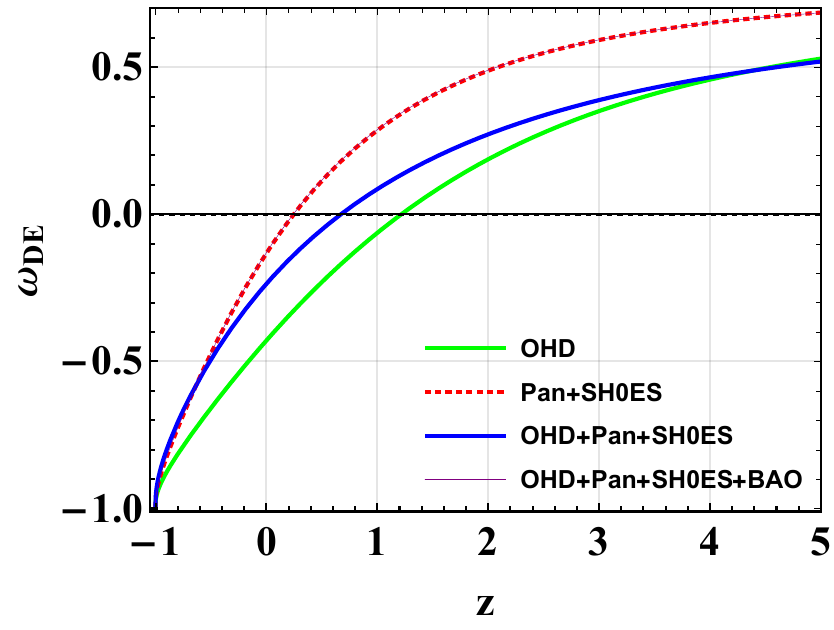}}
 \caption{ The evolution of the energy density ($ \rho $), the isotropic pressure ($ p $), and the EoS over the redshift ($ z $).}
\label{fig:rho-p}	
\end{figure}

In Fig. \ref{j}, the evolution of the jerk parameter $ j $ is observed. In this model, the trajectories of the jerk parameter are more consistent for the joint dataset  OHD + Pantheon + SH0ES, and for the other dataset, it deviates from the $\Lambda$CDM on high redshift $ z $. However, the trajectories of the jerk parameter converge to the $ \Lambda $CDM where $ j=1 $ as redshift $ z\to-1 $. In this model, we use the specific form of $ f(Q, C) $ function given in the Sec. \ref{sec:2}. The energy density $ \rho $ and the isotropic pressure $ p $ are evaluated from the field equations (\ref{17}-\ref{18}) and the Eqs. (\ref{20}-\ref{21}). Since the expressions of the energy density and the isotropic pressure are quite large, their evolutions have been studied using the depicted plots. In Fig. \ref{fig:rho-p}, the energy density and pressure decrease monotonically from the evolution of the early Universe to later times. The value of $ \rho $ is positive for the entire redshift range $z$, whereas $ p $ becomes negative in the redshift range $z<1.5$, and can be seen in Figs. (\ref{rh}-\ref{p}).  

The energy density $ \rho $, the pressure $ p $, and the EoS parameter $ \omega $ of the model play a crucial role in cosmology to understand the physical behavior of the Universe. The equation of state is given by the ratio between pressure and density, whose formula is $ \omega = p/ \rho $. In literature, the equation of state (EoS) parameter tracks the evolution of the Universe through various stages, which begin from an early deceleration stage to the late-times acceleration stage. These stages encompass periods such as the stiff fluid epoch ($\omega = 1$), radiation epoch ($\omega = \frac{1}{3}$), dust epoch ($\omega = 0$), quintessence dark energy ($-1 < \omega < -\frac{1}{3}$), the $ \Lambda $CDM ($\omega = -1$), and the phantom dark energy ($\omega < -1$). The quintessence dark energy represents a stable evolution of the Universe whereas the phantom dark energy is highly unstable.

In this model, the behavior of EoS can be analyzed in Fig. \ref{w}, where we can see different eras that our Universe passes through, depending on the redshift. At the present time, our model is in the quintessence region, and in late times, the value of $\omega$ converges to $-1$, converging to the same behavior of the $ \Lambda $CDM model.

\subsection{Energy Conditions}

\qquad The Energy conditions (ECs) are the essential features of the model to understand the Universe on a large scale, once the energy limits can explain the existence and formation of the matter. They help to characterize the ordinary and exotic matter distributions and can confirm the validity of the theoretical framework.

These energy conditions have been deduced from the considerations of the Raychaudhuri equation which explains the dynamics of the expansion of the late-time evolution of the Universe. It is remarkable to note that these equations are feasible to discuss in physical, geometrical, and perfect fluid form according to Table \ref{tab:Energy Conditions} and do not explicitly reference gravitational field equations. The criterion for gravitational attraction simplifies to $R_{ij} k^i k^j \geq 0$. In general relativity, this criterion is expressed in terms of the stress-energy tensor $T_{ij}$ as $T_{ij} k^i k^j \geq 0$. In modified gravity theories, the field equations take the form $G_{ij} \equiv R_{ij} - \frac{1}{2} R g_{ij} = T_{\text{eff}, \mu\nu}$. Consequently, the condition $R_{ij} k^i k^j \geq 0$ leads to the null energy condition (NEC) in the modified context as $T_{\text{eff}, ij} k^i k^j \geq 0$ \cite{Mandal/2020}. 
 
The Raychaudhuri equation holds universally across gravitational theories, it maintains its physical significance regarding the focusing of geodesic congruences and the attractive nature of gravity  \cite{Santos:2005pe, Santos:2007zza, Sen:2007ep, Qiu:2007fd}. Thus, the four types of energy conditions have been discussed in the modified gravitational field equations as given in Table \ref{tab:Energy Conditions} \cite{Visser:1997qk, Rani:2024uah}:

\begin{table}[htbp]
\caption{ \textbf{Energy Conditions}}
\centering
\begin{tabular}{l c c c c r}
\hline\hline
\\
Energy condition & \qquad Physical form & \qquad Geometric form & \qquad Perfect fluid form
\\
 \\
\hline
\\
 ~~  NEC\footnote{Null Energy Condition} ~~ & ~~ $ T_{ij}k^ik^j\geq 0 $ & ~~$ R_{ij}k^ik^j\geq 0 $  &  ~~ $  \rho_{\text{DE}}+ p_{\text{DE}} \geq 0 $
\\
\\ 
~~  WEC\footnote{Weak Energy Condition}  ~~ & ~~ $ T_{ij}t^it^j\geq 0 $  & ~~$  G_{ij}t^it^j\geq 0 $  & ~~$  \rho_{\text{DE}} \geq 0, \rho_{\text{DE}}+ p_{\text{DE}} \geq 0 $  
 \\
\\
~~  SEC\footnote{Strong Energy Condition}  ~~ & ~~  $ (T_{ij}-\frac{T}{n-2}g_{ij})t^it^j\geq 0  $ & ~~ $ R_{ij}t^it^j\geq 0 $  & ~~  $ \rho_{\text{DE}}+p_{\text{DE}} \geq 0, (n-3)\rho_{\text{DE}}+(n-1) p_{\text{DE}} \geq 0 $
\\
\\
~~  DEC\footnote{Dominant Energy Condition}~~  & ~~ $ T_{ij}t^i\xi^j\geq 0 $  & ~~ $ G_{ij}t^i\xi^j\geq 0 $ & ~~   $\rho_{\text{DE}} \geq | p_{\text{DE}}| $  
\\
 \\
\hline\hline
\end{tabular}
\label{tab:Energy Conditions}
\end{table}

\begin{figure}\centering
	\subfloat[]{\label{n}\includegraphics[scale=0.62]{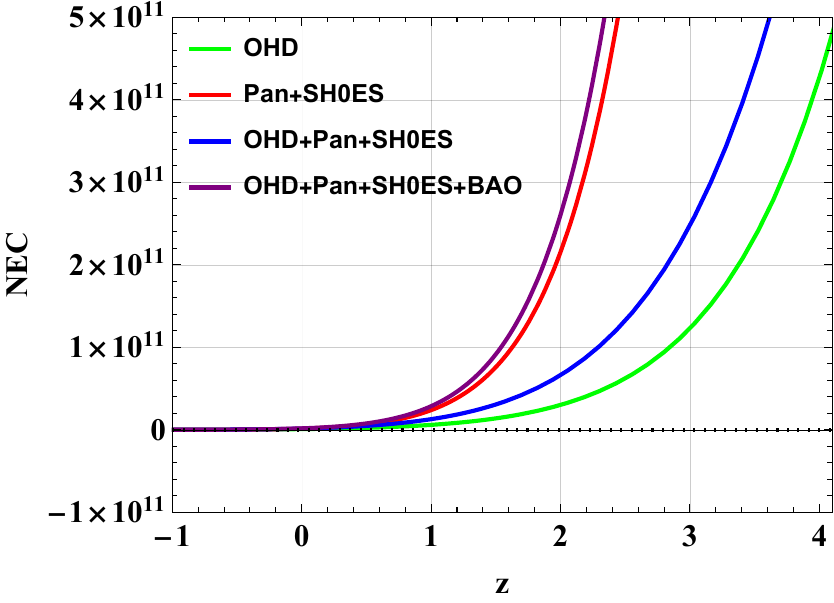}}	\hfill
	\subfloat[]{\label{s}\includegraphics[scale=0.64]{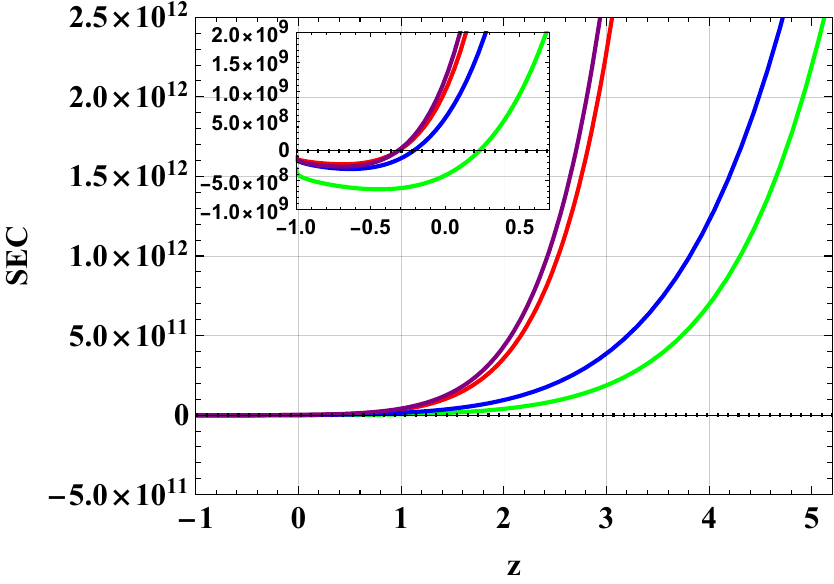}}\par
	\subfloat[]{\label{d}\includegraphics[scale=0.60]{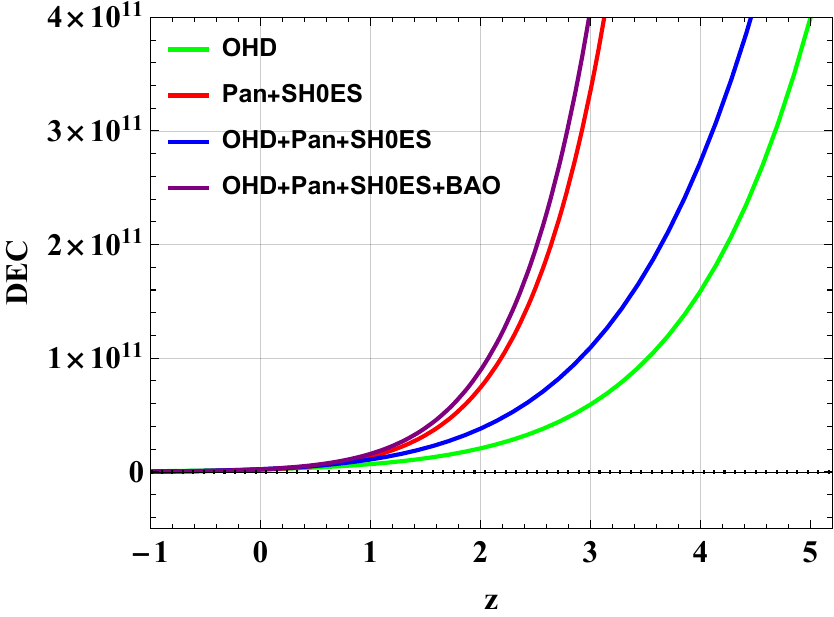}}
 \caption{ The dynamics of the NEC, SEC, and DEC over the redshift ($ z $).}
 \label{fig:5}
\end{figure}

In this model, the energy conditions NEC and DEC are satisfied for the entire redshift range $ z $ for all observational datasets but SEC violates for a redshift range of $ z<0.5 $ (see Fig. \ref{fig:5}). The violation of SEC indicates the presence of dark sectors in the Universe. Visser's theoretical prediction based on the energy conditions suggests that a negative cosmological constant might explain the violation of the SEC in the late-times evolution \cite{Visser:1997qk, Visser:1997tq}. Particularly, the violation of ECs does not necessarily mean a breach of the NEC. Even with exotic dark energy or fundamental mechanisms, violating the SEC does not automatically violate the NEC. Fig. \ref{fig:5} shows that both the NEC and the DEC hold. In contrast, the SEC is consistently violated across various best-fit values of the model parameters listed in Table \ref{tab1}. This violation implies the Universe's accelerated expansion in the distant future \cite{Singh:2024ckh, Singh:2019fpr, Bolotin:2015dja, Visser:1997qk, Visser:1997tq, Singh:2023bjx}

\subsection{Statefinder Diagnostic Technique}

\qquad The statefinder diagnostic is a useful tool to distinguish among the dark energy models. This geometric diagnostic tool was developed by Sahni et al. \cite{Sahni:2002fz, Sahni:2008xx} and denoted by the pairs $ \{r, s\} $ and $ \{r, q\} $, where $ r $, $ s $ are the statefinder parameters respectively. So, the statefinder parameters depend on the expansion dynamics of the Universe through high derivatives of the scale factor and are naturally one step beyond the Hubble parameter. The statefinder pair $ \{r, s\} $ is defined as 

\begin{equation}\label{37}
r = \frac{\overset{\cdot \cdot \cdot}{a}}{aH^3}, \quad s = \frac{r-1}{3\left(q - \frac{1}{2}\right)},
\end{equation}

\begin{figure}\centering
	\subfloat[]{\label{sr}\includegraphics[scale=0.505]{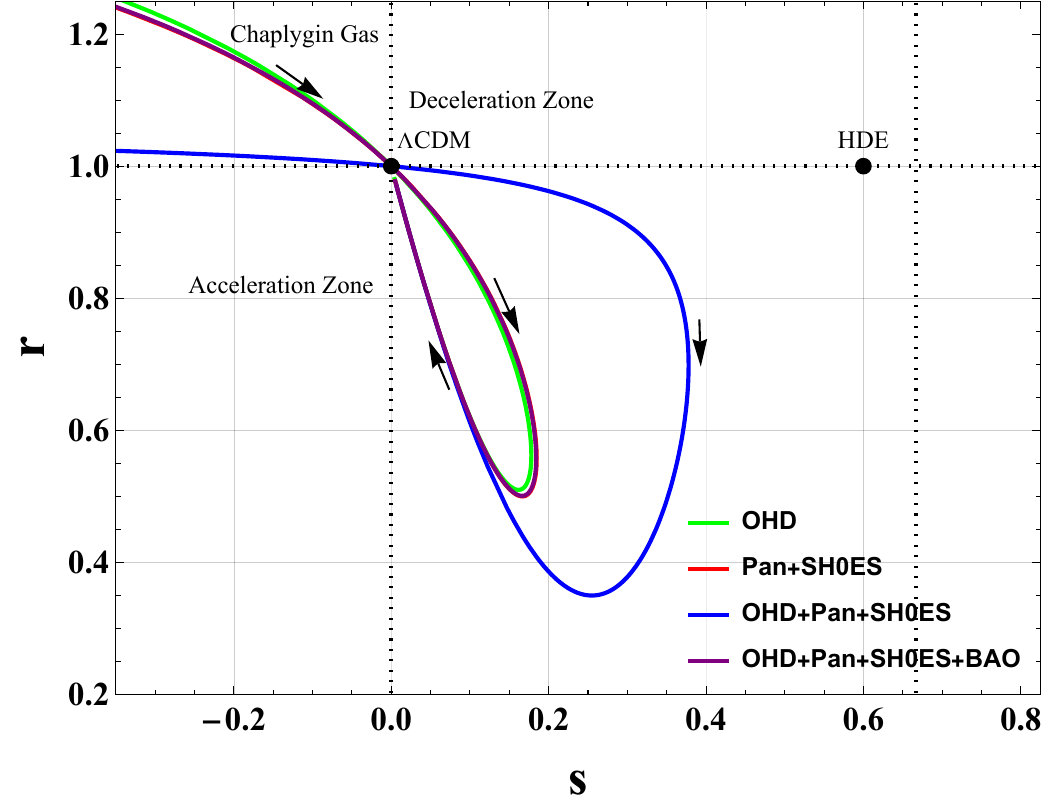}}\hfill
	\subfloat[]{\label{qr}\includegraphics[scale=0.505]{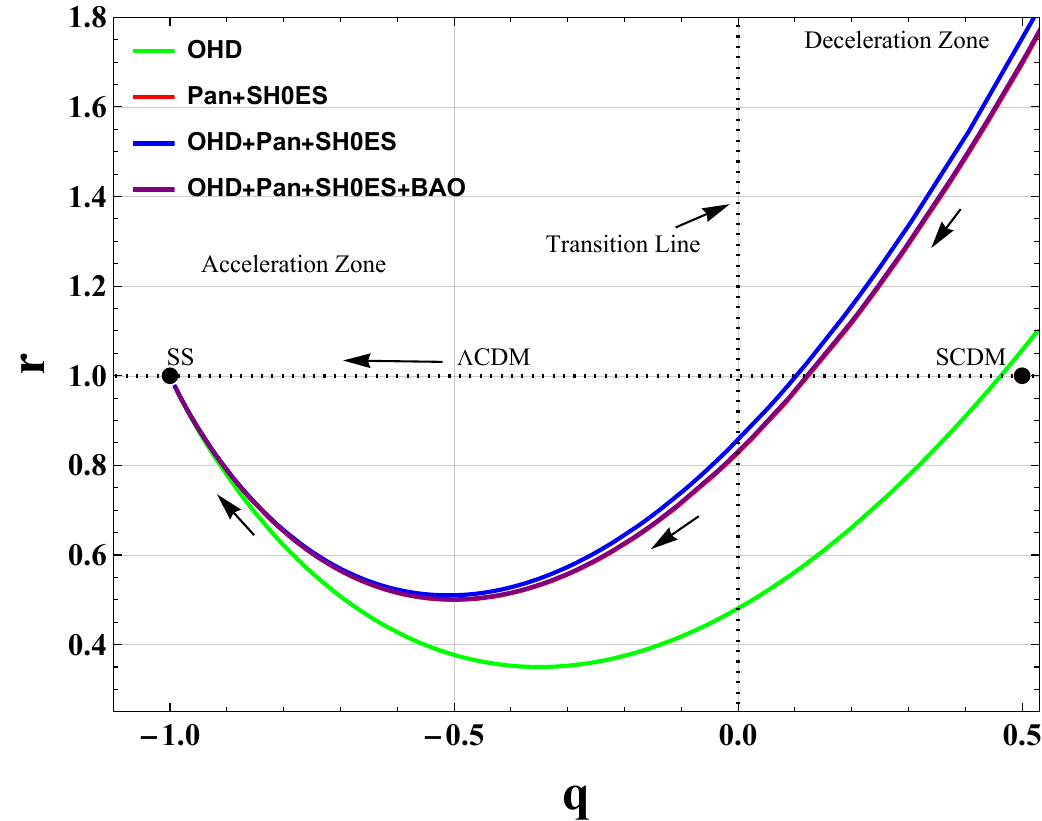}} 
\caption{ The parametric plots of the $ s-r $ and $ q-r $.}
\label{fig:6}
\end{figure}

In the $ s-r $ plane, the $\Lambda$CDM, HDE, and SCDM are denoted by the pairs $(s, r) = (0, 1)$, $(s, r) = (\frac{2}{3}, 1)$, and $ (s, r) = (1, 1) $ respectively. In the $ q-r $ plane, the $\Lambda$CDM is indicated by the line $ r=1 $ which is parallel to the q-axis. In this plane, the points SS and SCDM are denoted by the pairs $ (q, r) = (-1, 1) $, $ (q, r) = (\frac{1}{2}, 1) $ respectively. This technique has been widely employed in cosmology to differentiate among various kinds of DE models in the distinct modified theories (see Fig. \ref{fig:6}).

\begin{equation}\label{38}
    r=1+\frac{-3 \alpha -3 \alpha  W\left(\frac{\beta  e^{\beta /\alpha } (z+1)^{-1/\alpha }}{\alpha }\right)+2}{\alpha ^2 \left(W\left(\frac{\beta  e^{\beta /\alpha } (z+1)^{-1/\alpha }}{\alpha }\right)+1\right)^3},
\end{equation}
and 
\begin{equation}\label{39}
    s=2 \left(\frac{1}{-6 \alpha -3 \alpha  W\left(\frac{\beta  e^{\beta /\alpha } (z+1)^{-1/\alpha }}{\alpha }\right)+\frac{2-3 \alpha }{W\left(\frac{\beta  e^{\beta /\alpha } (z+1)^{-1/\alpha }}{\alpha }\right)}}+\frac{1}{3 \alpha +3 \alpha  W\left(\frac{\beta  e^{\beta /\alpha } (z+1)^{-1/\alpha }}{\alpha }\right)}\right).
\end{equation}
The left and right panels of Fig. \ref{fig:6} depict the trajectories of our model across various observational datasets in both the $ s-r $ and $ q-r $ planes. The evolution of the $ s-r $ trajectories traverse via the Chaplygin gas region and the $\Lambda$CDM and enters into the quintessence region before converging towards the $\Lambda$CDM in the later epochs which can be shown in Fig. \ref{sr}. In Fig. \ref{qr}, the trajectories deviate from the SCDM before converging to the point $ (-1, 1) $ known as the steady-state model (SS) in the $ q-r$ plane in the later times.

\subsection{Perturbation Analysis}
\qquad In the context of cosmic evolution, even minute variations in the density of a uniform and isotropic fluid play a crucial role in shaping the formation of large-scale structures of the Universe. These variations arise because the influence of pressure is relatively minor compared to the dominant gravitational forces. Consequently, the colossal cosmic structures observed today, such as galaxies and galaxy clusters, trace their origins back to these initial fluctuations \cite{Fry:1983cj, Bharadwaj:1996qm, delaCruz-Dombriz:2011oii, Bhardwaj:2022lrm, Jaybhaye:2022gxq, Narawade:2022jeg}. To assess the stability of our theoretical framework under $f(Q, C)$ gravity, we specifically examine linear perturbations. These perturbations manifest as fluctuations in both the Hubble parameter and energy density.

\begin{equation}\label{40}
	H(t) = H_0(t)(1+\delta_H(t)) ~~\text{and}~~\rho(t) = \rho_0(t)(1+\delta_\rho(t)),
\end{equation}
where, $ H_0(t) $, $ \rho_0(t) $, $ \delta_H(t) $, and $ \delta_\rho(t) $ stand for the baseline Hubble parameter, baseline energy density, and perturbation parameters for the Hubble parameter and energy density, respectively. 

Within the domain of matter fields, we confront equations of first-order perturbations. The particular equation employed in this model is
\begin{equation}\label{41}
	\dot{\delta_\rho}(t) + 3H_0(t)\delta_H(t) = 0.
\end{equation}

The analytical connection between perturbations in matter and geometry is established as
\begin{equation}\label{42}
b \delta_\rho(t) = -6\left[{H_0(t)}\right]^2\delta_H(t),
\end{equation}
where $ b = K\rho_{m0} $ which is normalized to unit in this work. Here, the fluctuations in matter dictate the total perturbation around a cosmological solution (or vice versa) in the context of General Relativity (GR). The first-order perturbation equation employed in this model can now be rewritten by taking $ \delta_H(t) $ in eq. (\ref{42}) and replacing it into (\ref{41}), yielding to
\begin{equation}\label{43}
\dot{\delta_\rho}(t)-\frac{b}{2H_0(t)}\delta_\rho(t) = 0.
\end{equation}
Then, by integrating Eq. (\ref{43}), we get
\begin{equation}\label{44}
\delta_\rho(t) = C~\exp\left[\frac{1}{2} \int \frac{b}{H_0(t)} dt\right].
\end{equation}
Also, the evolution of the Hubble perturbation $ \delta_H(t) $ is given as
\begin{equation}\label{45}
\delta_H(t) = -\frac{b}{6\left[{H_0(t)}\right]^2} C_1~\exp\left[\frac{1}{2} \int \frac{b}{H_0(t)} dt\right].
\end{equation} 
Thus, the functions $ \delta_\rho(t) $ and $ \delta_H(t) $ can be expressed in terms of redshift $ z $ as
\begin{equation}\label{46}
\delta_\rho(z) = C~\exp\left[\frac{1}{2} \int \frac{b}{(1+z)~\left[{H_0(z)}\right]^2} dz\right]
\end{equation}
and
\begin{equation}\label{35}
\delta_H(z) = -\frac{b}{6\left[{H_0(z)}\right]^2} C_1~\exp\left[\frac{1}{2} \int \frac{b}{(1+z)~\left[{H_0(z)}\right]^2} dz\right].
\end{equation}
Here, $ C $ and $ C_1 $ are the arbitrary integration constants and for further analysis, we take $C=1$ and $C_1=-1$.

\begin{figure}\centering
	\subfloat[]{\label{ph}\includegraphics[scale=0.51]{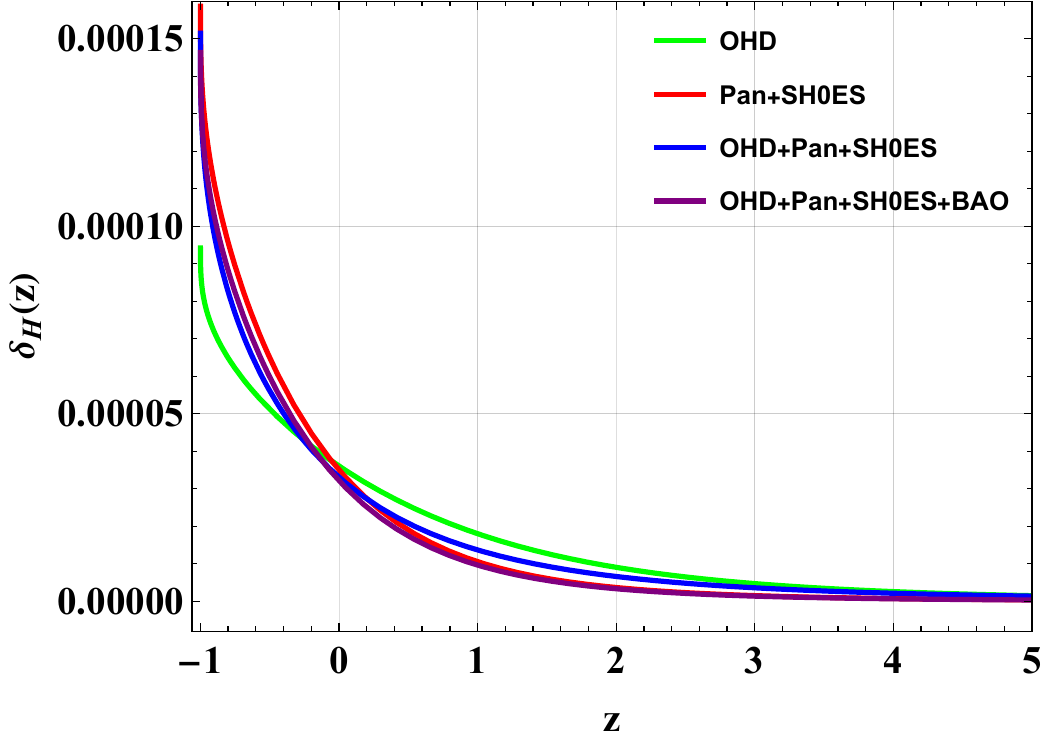}}\hfill
	\subfloat[]{\label{prh}\includegraphics[scale=0.5]{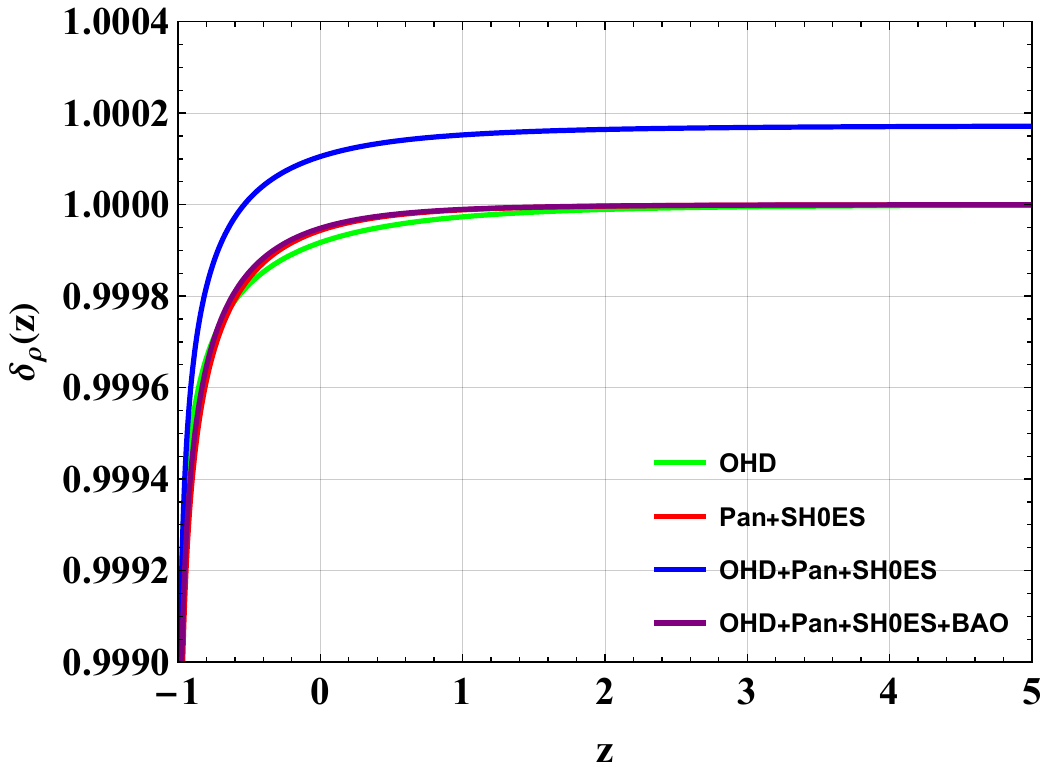}} 
\caption{ The evolution of the perturbations in the energy density $ \delta_\rho(z) $ and in the Hubble parameter $ \delta_H(z) $ over the redshift $ z $.}
\label{perturbation}
\end{figure}
	
To study the perturbation parameters, we have plotted the trajectories of $\delta_\rho(z)$ and $\delta_H(z)$ for all the observational datasets. In Fig. \ref{perturbation}, we notice that the variation of perturbation parameters attain their small values throughout the whole range of $z$ and this indicates the stability of the model.

\section{Conclusions} \label{sec:5}
\qquad In the symmetric teleparallel approach to gravity, non-metricity takes precedence over curvature as the fundamental geometric quantity. Unlike the Levi-Civita Ricci scalar $R^*$, the non-metricity scalar $ Q $ differs by an additional term $ C $ that represents a total divergence term as $ R^*=Q+C $. At the level of equations, general relativity and symmetric teleparallel general relativity are equivalent. However, modifications involving functions of $ R^* $ and $ Q $ are not equivalent due to their distinct differential properties. In our study, we have studied $ f(Q, C) $ gravity and cosmology by incorporating both $ Q $ and the boundary term $ C $ into the Lagrangian formulation. Initially, we derived the general field equations and subsequently applied them to the flat Friedmann-Robertson-Walker (FRW) metric within a cosmological context. By exploring connection choices, our analysis uncovered an effective dark-energy component stemming from geometric considerations, which yields intriguing cosmological phenomena. Furthermore, our framework demonstrated the potential for an effective interaction between the matter and the dark energy.

Using a specific form of the function $ f(Q, C) $, we have illustrated how our model mirrors the conventional thermal history of the Universe, encompassing epochs dominated by both matter and dark energy as required. Moreover, we have parameterized the scale factor and examined the behavior of dynamical and physical cosmic parameters by calculating the best-fit values of the model parameters. To obtain the optimal values of the model parameters $ \alpha $, $ \beta $, and the Hubble parameter $ H_0 $, we use the Markov Chain Monte Carlo (MCMC) method by applying the emcee codes in Python using various observational datasets such as OHD, Pantheon, SH0ES, BAO. The constraints and features are shown in detail in Figs. \ref{contour1}-\ref{contour4}. The posterior distribution of the reconstructed deceptive $ f (Q, C) $ model under $ 1\sigma $ and $ 2\sigma $ CLs and the error bar plots of the Hubble parameter $ H(z) $, distance modulus $ \mu(z) $ exhibit the degree of consistency between our model and the $ \Lambda $CDM (see Figs. \ref{errorbar}). 
 
The current values of the physical parameters  $ H $  and $ q $ and the model parameters for different observational datasets are accessed in Table \ref{tab1}. We investigated the behavior of the trajectories of the deceleration parameter $ q $ for the best-fit values in Fig. \ref{q} where the transition from the deceleration to acceleration occurs in the redshift range  $  0.70 < z_{tr} < 2  $. The deceleration parameter also corroborates with an accelerating expansion of our Universe at the present time. The jerk parameter $ j $ is positive for all the redshift ranges and converges to $ 1 $ as $ z\to-1 $ for all the constrained model parameters shown in Fig. \ref{j}. The evolution of the energy density is always positive and decreases as the cosmic redshift $ z\to-1 $ (see Fig. \ref{rh}). 

The isotropic pressure of the model is positive for the redshift range $ z>1 $ and shows the negative behavior for all the constraints of the model parameters as $ z\to-1 $. The model indicates the existence of dark energy in the Universe that leads to the accelerating expansion of the Universe in later times (see Fig. \ref{p}). The EoS parameter transits from the perfect fluid state to the dark energy state, approaching $ -1 $, as shown in Fig. \ref{w}. Hence, the model unveils the quintessence-like dark energy model and corroborates with $ \Lambda $CDM as $ z\to-1 $. In fig. \ref{fig:5}, we can verify that the model satisfies the energy conditions NEC and DEC but does not satisfy SEC for later times, indicating the existence of a dark energy regime.

Fig. \ref{fig:6} depicts the behaviors of the statefinder diagnostic pairs $ \{s, r\} $ and $ \{q, r\} $. By analyzing the parametric space $ \{s, r\}$, we observe that the model starts in a Chaplygin region, enters the quintessence region, and then corroborates with $ \Lambda $CDM. Therefore, the model transits from the deceleration to the acceleration regime, deviates from SCDM, crosses the Quintom line, and approaches the de-Sitter state in the parametric space $ \{q, r\} $. In Fig. \ref{perturbation}, we have depicted the trajectories of $\delta_\rho(z)$ and $\delta_H(z)$ for all the observational datasets. It has been noticed that the variation of perturbation parameters attain their small values throughout the whole range of redshift $ z $, which indicates the stability in the model.

Finally, given the evolution of the various trajectories of the jerk parameter, the EoS parameter, and the statefinder, we can see that our model corroborates with the $ \Lambda $CDM. Moreover, in analyzing the stability of the model, we discuss the perturbations for both the Hubble parameter and the energy density. Hence, we conclude that the model reflects an accelerated expanding model with quintessence-like behavior at later times $ z\to-1 $. 

It is relevant to point out that the techniques presented in this work can be applied to other models of gravity, such as in $f(R,T) - \Lambda(\phi)$ \cite{Santos/2023}, $f(T, {\cal T})$ \cite{Harko/2014}, $f(Q,T)$ \cite{Arora/2021}, and Weyl $f(Q,T)$ \cite{Xu/2020}. Moreover, we can also investigate in detail the mapping of further data in low redshift ranges such as those from DESI \cite{DESI/2024} and also from new and future surveys like Euclid \cite{Euclid/2024} and BINGO \cite{BINGO/2022, BINGO-II/2022} telescopes.

\vskip0.2in

\section*{Acknowledgement} The authors J. K. Singh and Shaily thank the Dept. of Mathematics, NSUT, New Delhi-78, India, and Bennett University, Greater Noida, India for providing the necessary facilities where a part of this work has been completed. JRLS would like to thank CNPq (Grant no. 309494/2021-4), and FAPESQ-PB (Grant 11356/2024) for financial support. 

\vskip0.2in

\section*{Data Availability Statement} In this manuscript, we have used observational data as available in the literature as such our work does not produce any form of new data.

\end{document}